\documentstyle[12pt]{article}
\begin{document}
\begin{center}
\section*{Saturation properties and liquid-gas phase transition
of nucleus\footnote{Supported  in  part  by National 
Natural Science Foundation of China.}}

\vspace{1.cm}

Fu-guang Cao$^{a,b}$ and Shan-de Yang$^{a}$\\

\vspace{0.5cm}
$a$ Center of Theoretical Physics and Department of Physics,
Jilin University, Changchun, 130023, P. R. China\\
$b$ Institute of Theoretical Physics, Academia Sinica,
P.O.Box 2735, Beijing, 100080, P. R. China\footnote{Mailing address.
E-mail address: caofg@itp.ac.cn.}\\

\end{center}

\vskip 0.5cm
\begin{center}
PACS Number(s): 21.60.Jz, 21.10.Dr, 25.70.Hi 
\end{center}
\vskip 2cm

\begin{abstract}

Saturation properties and liquid-gas phase transition
of nucleus are analysed
in the framework of Hatree-Fock theory.
We modify Hill-Wheller formula with a
finite-size-effect parameter
by fitting the zero-temperature properties of nucleus.
Employing Gogny effective interaction and phenomenological expression
of Coulomb energy,
we give the critical
temperature of liquid-gas phase transition of
nucleus being about $12$ MeV, which agrees with the result 
extracted  from heavy-ion collision experiments.  
It is pointed out that a
phenomenological formula of surface energy of hot nucleus
is not avaliable in the region where nucleon density is far away
from the normal density.

\end{abstract} 


\newpage
\noindent
{\bf I. Introduction} 

The developments of heavy-ion physics have evoked great interest on
the studies about statistic properties of infinite nuclear matter 
and nucleus, especially about equation of state (EOS) 
and liquid-gas phase transition.
There have been a lot of calculations of infinite nuclear matter
in the framework of Hatree-Fock theory(HFT), in which finite-size-effect 
and Coulomb interaction were ignored [1-10].
The finite-size-effect of nucleus and Coulomb interaction
in nucleons
should play important roles in realistic heavy-ion collisions.
Refs. \cite{Jaqaman,Song} took into account the finite-size-effect with
the help of Hill-Wheller formula and Coulomb effect in the use of
a phenomenological expression in the calculation of
finite nuclear matter.
It was found that 
the finite-size-effect leads to a reduction of
critical temperature of liquid-gas transition in finite
nuclear matter by about $5\sim 10$ MeV, 
and Coulomb effect leads to a reduction
about $1\sim 3$ MeV.
In our opinion, any approach which is applied to study
hot nucleus though infinite nuclear matter approximation
should be able to give reasonable zero-temperature 
saturation properties of nucleus also.
Basing on the calculations of six typical 
nuclei (from $^{40}_{20}Ca$ to $^{238}_{92}U$) with several Skyrme 
effective interactions, Wang and Yang \cite{Wang}
found that it is impossible to give correct 
bound energies
and other saturation properties of nucleus
by adopting Hill-Wheller formula directly as that 
in Refs. \cite{Jaqaman,Song}.
There is a general agreement that Gogny 
effective interaction \cite{Gogny} is able to
describe the long-range and medium-range behaviours of nucleon-nucleon 
interaction more reasonable than Skyrme interaction.
In this paper, we will modified Hill-Wheller formula with a 
finite-size-effect
parameter through the studies on the zero-temperature saturation
properties of nucleus with Gogny interaction.
Employing the modified Hill-Wheller formula, we study the liquid-gas
transition of hot nucleus.
We also analyse the applicability of a widely used phenomenological
expression of surface energy of hot nucleus.

The paper is organized as follows.
In section II, we 
modified Hill-Wheller formula by fitting
the zero-temperature properties of nucleus.
We study the stability character and liquid-gas phase transition of
hot nucleus in section III.
The surface energy 
of nucleus is discussed in section IV.
As usual, the last section is reserved for summary.

\vspace{0.5cm}
\noindent
{\bf II. Formalism}

Let us begin with a brief review of Hatree-Fock theory of 
infinite nuclear matter system with Gogny interaction.
Gogny D1 effective interaction is expressed as \cite{Gogny}
\begin{eqnarray}
V({\bf r}) &=& \sum_{i=1,2}\left(W_{i}+B_{i}P_{\sigma}-H_{i}P_{\tau}
-M_{i}P_{\sigma}P_{\tau}\right) \exp(-r^{2}/{\mu}^{2}_{i})\nonumber\\
 & & +t_{0}\left(1+X_{0}P_{\sigma}\right) \rho^{\alpha}
\delta({\bf r})
+iW_{LS}({\bf \sigma}_1+{\bf \sigma}_2)\cdot\bigtriangledown\times
\delta({\bf r}) \bigtriangledown,
\label{Gogny}
\end{eqnarray}
where $\rho$ is nucleon density and $P_{\sigma},P_{\tau}$ are 
exchange  operators of spin and isospin respectively. 
In Eq. (\ref{Gogny}), the part 
with $\delta$ function is zero-range interaction which is
density-dependent and the part with Gauss function is 
finite-range interaction which is density-independent. 
The parameters of Gogny
D1 interaction are listed in table~1.
The Hatree-Fock single particle spectrum with Gogny D1
interaction is 
\begin{eqnarray}
\epsilon_{q} &=& \frac{h^{2}q^{2}}{2m}+u_{q}+u_{D}, \nonumber \\
u_{q} &=& \sum_{q'}\left(qq'\left|V\right|qq'-q'q\right)f_{q'}, \nonumber \\
u_{D} &=& \frac{1}{2V}\sum_{i\not= j}\left(q_{i}q_{j}\left|
\frac{\partial V}
{\partial \rho}\right|q_{i}q_{j}-q_{j}q_{i}\right)
f_{q_{i}}f_{q_{j}}, \nonumber \\
f_{q} &=& \frac{1}{1+exp\left[\beta(t_{q}+u_{q}+u_{D}
-\mu)\right]},
\label{sp}
\end{eqnarray}
where  $V$ is the volume of nuclear matter system,
$\mu$ the chemical potential, 
$T$ the temperature, $\beta=1/(k_{B}T)$, 
$u_{q}$  the  usual  HF single particle   potential,  and  $u_{D}$
the rearrangement potential. It has been pointed out that
the introduction  of  rearrangement potential is necessary to guarantee
the density relation
$\rho=\frac{1}{V}\sum_{q}f_{q}$
and give correct chemical  potential \cite{Yang}.
Then EOS can be readily obtained from Eq. (\ref{sp}) 
\cite{Heyer,Song,Huang}.

Now we turn to the approach to respect finite-size-effect
and Coulomb interaction in finite nuclear matter system.
In Hatree-Fock theory, the calculation of infinite nuclear
matter is simplified by taking plane wave function as
single particle wave function, in which the state number between 
$k$ and $k+dk$ in the momentum space reads
\begin{eqnarray}
dN_{k}=V\frac{k^{2}dk}{2\pi^{2}}.
\label{wni}
\end{eqnarray}
To study the liquid-gas phase transition 
in finite nuclear 
matter system, Jaqaman, Mekjian and Zamick \cite{Jaqaman} 
proposed that the finite-size-effect can be taken into account 
in the theory framework of infinite nuclear matter system 
by Hill-Wheller formula
\begin{eqnarray}
dN_{k}=V\left[\frac{k^{2}dk}{2\pi^{2}}-\frac{S}{V}\frac{kdk}{8\pi}
+\frac{L}{V}\frac{dk}{8\pi}\right],
\label{wnj}
\end{eqnarray}
where S and L are the measures of the average surface and linear 
of finite nuclear matter system
respectively. For a spherical system with radius $R$, $S=4\pi R^{2}$,
$L=2\pi R$. 
The effects of zero-point motion
and quantization of wave number in a finite-size system
were taken into account in Eq. (\ref{wnj}).
In fact, Eq. (\ref{wnj}) is derived from a simple model, a finite-size
system with free particles.
Strictly speaking, it cannot be applied to the
study of nucleus directly.
The reason is as following: 
Because there are interactions
in nucleons, the single particle wave function in nucleus is not 
plane wave function as that in the finite-size
system without interaction,
and different for the 
nucleus with different shell.
Eq. (\ref{wnj}) is a approximate approach to study
finite-size-effect in the theory framework of nuclear
matter system. 
In our opinion, any approach which is applied to 
study the liquid-gas phase 
transition of nucleus through the infinite nuclear
matter approximation should be able to give reasonable 
zero-temperature saturation properties of nucleus also.
The calculations of the six typical nuclei show that
Eq. (\ref{wnj}) can not satisfy this constrain (see table 2).
It can be found that 
the bound energies per nucleon of the nuclei 
calculated with Eq. (\ref{wnj}) are  much  smaller 
than the experimental data and this divergence becomes more and more 
serious as the nucleon number decrease, so that $^{40}_{20}Ca$
can not combine any more.
This shows that the finite-size-effect isn't 
considered suitably with Eq. (\ref{wnj}). 
As a result, it is hardly to believe the calculations 
of EOS and liquid-gas transition of finite nuclear matter system
given in Refs. \cite{Jaqaman,Song}.
We may add a finite-size-effect parameter $a_F$
to the second and third
terms in Eq. (\ref{wnj}) to respect the fact
that the single particle wave function in
nucleus is different from that in the finite
system with free particle,
\begin{eqnarray}
dN_{k}=V\left\{\frac{k^{2}dk}{2\pi^{2}}+a_{F}\left[
-\left(\frac{4\pi\rho}
{3A}\right)^{\frac{1}{3}}\frac{3kdk}{8\pi}+
\left(\frac{4\pi\rho}{3A}\right)
^{\frac{2}{3}}\frac{3dk}{16\pi}\right]\right\},
\label{wncao}
\end{eqnarray}
where  $a_{F}$ is decided by fitting the experimental data of bound 
state energies of nuclei.
Eq. (\ref{wncao}) becomes Eq. (\ref{wnj}) when we take $a_F=1$
and becomes Eq. (\ref{wni}) as $A\rightarrow\infty$.
By fitting  the  experimental data,
we find that to take the finite-size-effect
parameter $a_{F}=0.35$ in Eq. (\ref{wncao}),  
the bound energies
of the nuclei from light to heavy 
agree  with  the experimental data
very well and 
the saturation densities $\rho_0$ of the six 
typical nuclei are about 0.13,  which 
has mild dependence on the nucleon number $A$ (see table 2). 
This  agrees 
with  the  saturation  character. With this $\rho_{0}$,
through the phenomenological formula 
$R=r_{0}A^{\frac{1}{3}}$ of  the effective radius 
$R=\left( 3/4\pi \rho_{0} \right)^{\frac{1}{3}}$,
we obtain $r_{0}=1.21\sim 1.22$, 
which  agrees with the experimental result also.
All of these show that 
it is reasonable with
Eq. (\ref{wncao}) to study the finite-size-effect
of nucleus.

Coulomb 
interaction is a long-range interaction, adopting it directly in 
the calculation of matrix between plane wave 
functions will bring divergent result. So, for simplicity, we
use a phenomenological expression of Coulomb energy per
proton to respect Coulomb effect \cite{Jaqaman,Song},
\begin{eqnarray}
E_{C}=a~Z^{2}\left[1-5\left(\frac{3}{16\pi Z}\right)^{\frac{2}{3}}-
\frac{1}{Z}\right]\rho^{\frac{1}{3}}A^{-\frac{4}{3}}.
\label{col}
\end {eqnarray}
We take $a=1.50$ in order to fit the experimental data 
more well,
which is slight different with 
the usual value $a=1.39$.

\vspace{0.5cm}
\noindent
{\bf III. Stability character and liquid-gas phase transition of 
hot nucleus} 

For discussing the stability character of hot nucleus, 
limit temperature $T_{l}$ is defined, which is the highest 
temperature below that free energy has a minimum.
Nucleon 
density corresponding with the minimum of free energy at $T_{l}$ is
called limit
density $\rho_{l}$. When the temperature is larger than  
$T_{l}$, the 
pressure of the system is always larger than zero. The
condensation phase and the nucleon-gas phase can not arrive  phase 
equilibrium without extra pressure. The values of $T_{l}$ and 
$\rho_{l}$ of the six typical nuclei are listed in table 3.
It can be found that $T_{l}$  and $\rho_{l}$ are 
not sensitive to nucleon number $A$.  
$T_{l}$ is about $9.0\sim 9.5$ MeV, 
and $\rho_{l}$ is about 0.075 fm$^{-3}$.

Liquid-gas phase transition of hot nucleus is another
interesting question. Strictly speaking, a phase transition 
can only occur in a system with infinite number of particles, 
which is reflected in the singularity behavior of some 
thermodynamic quantities. For example, the specific heat displays 
a sharp $\lambda$-type singularity at the critical temperature
$T_{c}(\infty)$ for
a liquid-gas phase transition in an infinite particles system.
The specific heat of
finite particles system does not exhibit such a sharp 
singularity, but it has a large peak at a temperature $T_{c}(A)$ which 
approaches to $T_{c}(\infty)$ as the particle number 
$A\rightarrow\infty$.
The temperature 
$T_{c}(A)$ can be regard as the critical temperature of a finite 
particle system.
For a finite particles system, it is convenient to determine the 
critical point by 
the inflection point condition of $\mu\sim\rho$ isothem 
rather than
$P\sim\rho$ isothem \cite{Jaqaman}
\begin{eqnarray}
\left. \frac{\partial\mu}{\partial\rho}\right| _{\gamma,T_{c},\rho_{c}}=
\left. \frac{\partial ^{2}\mu}{\partial\rho^{2}}
\right| _{\gamma,T_{c},\rho_{c}}=0,
\end{eqnarray}
where $\mu$ is the average chemical potential
\begin{eqnarray}
\mu=\frac{1}{\rho}(\rho_{p}\mu_{p}+\rho_{n}\mu_{n})=
\frac{1-\gamma}{2}\mu_{p}+\frac{1+\gamma}{2}\mu_{n}
\end{eqnarray}
with $p$ denotes parton and $n$ neutron.
To study the finite-size-effect, we calculate the EOS and
liquid-gas transition critical point of 
symmetrical infinite nuclear matter system with different nucleon number
and without Coulomb interaction.
Figs. 1 and 2 display $\mu\sim \rho$ isotherms at temperature 
$T=6$ MeV and  $14$ MeV. As it is known, liquid 
phase and gas phase can coexist at low temperature,
while only gas phase can exist at high temperature.
It can be found that there is a critical nucleon number $A_{c}(T)$
for a certain temperature.
As $A\leq A_{c}(T)$, only gas phase can exist. Table 4 lists 
the critical temperatures $T_{c}$ and densities $\rho_{c}$
of the symmetrical nuclear matter 
with different nucleon number $A$. $T_{c}$ and $\rho_{c}$ decrease 
with $A$ decreasing.  Here again we find that 
for the realistic nucleus  system 
$(A<250)$ the liquid-gas critical points calculated with 
Eq. (\ref{wncao}) and $a_{F}=0.35$ are very different from that with 
Eq. (\ref{wnj}) 
obviously.
Coulomb interaction and neutron-proton asymmetry are 
another two factors which decrease the critical temperature. 
Table 3 gives the critical temperatures $T_{c}$ and
densities $\rho_{c}$ of the six typical nuclei.
It can be found that
with Eq. (\ref{wncao}) and $a_{F}=0.35$, 
finite-size-effect, Coulomb interaction
and asymmetry effect reduce the critical 
temperature $T_{c}$ by about $3\sim 4$ MeV 
comparing with that of the symmetrical infinite nuclear matter.
$T_{c}$ and $\rho_{c}$ are
insensitive to nucleon number $A$, which is very different from the 
results with Eq. (\ref{wnj}). 
As exhibited in section II, we can 
obtain reasonable zero-temperature saturation properties of
nuclei with Eq. (\ref{wncao}) and $a_{F}=0.35$ while we cannot 
with Eq. (\ref{wnj}).
Thus the  critical  temperatures $T_{c}$ obtained in the 
present work are reasonable and reliable. 
$T_c$ has mild dependence
on nucleon number $A$.
In fact, all of the finite-size-effect, Coulomb 
interaction and neutron-proton asymmetry play the  roles 
to weaken the nucleus combination and reduce $T_{c}$. 
The  finite-size effect becomes weak
with $A$ increasing, while Coulomb 
interaction and neutron-proton asymmetry effect play more 
and more important roles.  
As a result, $T_c$ is not sensitive to the nucleon number $A$, 
$T_c \simeq 12$ MeV.
Our results are close to 
that with Skyrme 
effective interaction \cite{Wang}. Panagiotou  {\em et.
al.} \cite{Pan} extracted the critical 
temperatures of nuclei by analyzing the mass distribution of 
multi-fragments
in intermediate-energy heavy-ion collisions basing on the 
condensation theory, which gave $T_{c} \simeq 12$ MeV.
Bondorf {\em et.al.} \cite{Bondorf} obtained $T_{c} \simeq 11$ MeV
for A=100 system basing on the Monte Carlo calculation.
Our calculations are consistent with these results.

\vspace{0.5cm}
\noindent
{\bf IV. Surface energy of nucleus}

A phenomenological surface energy expression
was applied to study the stable properties and dynamical 
properties of nucleus
in some works \cite{Jaqaman2,Band},
\begin{eqnarray}
E_{S}=\frac{4\pi R^{2}\sigma}{A}=4\pi\left(\frac{3A}{4\pi\rho}\right)
^{\frac{2}{3}}\frac{\sigma}{A},
\label{sfp}
\end{eqnarray}
where $\sigma$ is the tension 
coefficient $\sigma \simeq 1.2$ MeV$\cdot$fm$^{-2}$ which is derived from 
the surface energy term in the 
bound energy phenomenological formula of nucleus.
This approach takes the tension
coefficient being 
independent of density, which gives the relation
$E_{S}\sim\rho^{-\frac{2}{3}}$.
In the present calculations, the surface energy coming
from the finite-size-effect can be derived naturally from
the bound energy of nucleus
and that of infinite symmetric nuclear matter, 
\begin{eqnarray}
\Delta E=E(A,\rho)-E(A\rightarrow\infty,\rho).
\label{sfc}
\end{eqnarray}
The surface energy
of $^{90}_{40}Zr$ obtained with Eqs. (\ref{sfp}) and (\ref{sfc}) are  
plotted in Fig. 3.
Although Eqs. (\ref{sfp}) and (\ref{sfc}) 
present similar results in the neighborhood of
normal nucleon density $\rho_{0}$,  
their behaviors are very different in the other density regions.
The saturation curves 
$(E/A)\sim \rho$, pressure-density curves $P\sim \rho$
calculated with Eq. (\ref{sfp}) are plotted in 
Figs. 4 and 5, which are very different form that obtained form HFT.
The saturation properties of nucleus calculated with Eq. (\ref{sfp}) 
are given table 2.
We find that according to Eq. (\ref{sfp}),
the bound energies agree with the experiment data, but the saturation 
densities of the nuclei are general larger than that of
infinite nuclear matter and reduce as mass number $A$
increasing, which is not reasonable obviously.
All of these show 
that Eq. (\ref{sfp}) is not available for the region 
where the nucleon density is far 
away from the normal nucleon density.
We find that Eq. (\ref{sfc}) can be expressed by the following 
formula
\begin{eqnarray}
E'_{S}=C_{1}\rho^{\frac{2}{3}}A^{-\frac{1}{3}}+C_{2}\rho^{\frac{5}{3}}
A^{-\frac{1}{3}},
\label{sfphcao}
\end{eqnarray}
where  $C_{1}=53.0$ MeV$\cdot$fm$^{2}$ and $C_{2}=108$ MeV$\cdot$fm$^{5}$.
In Eq. (\ref{sfphcao}) the first term is derived from the fermion gas model
and the finite-size-effect is taken into account by the second term.
The density dependence of Eq. (\ref{sfphcao}) 
which reads 
$E'_{S}\sim\rho^{\frac{2}{3}}$
is very different form
that of Eq. (\ref{sfp}).
Tables 2 and 4 give the saturation  properties 
and critical temperature in the use of Eq. (\ref{sfphcao}),
which agree with the results with Eq. (\ref{wncao}) very well.
The calculations also show that the surface energy Eq. (\ref{sfc})
is not sensitive to 
temperature. Thus, Eq. (\ref{sfphcao}) 
can be employed in the study about EOS of nucleus.

\vspace{0.5cm}
\noindent
{\bf V. Summary and discussion}

We analyse the saturation properties and
liquid-gas phase transition of nucleus and infinite nuclear
matter in the framework of Hatree-Fock theory.
At first, 
in order to give reasonable zero-temperature properties of
nucleus,
we modify Hill-Wheller formula
which is used to respect the finite size effect of nucleus.
Employing Gogny effective interaction and phenomenological 
Coulomb energy,
we obtain the critical temperatures of liquid-gas phase transition
of nucleus being about 12 MeV, which is consistent with the
result extracted from heavy-ion collisions. 
The critical temperature of
liquid-gas phase transition of nucleus has mild dependence
on nucleon number $A$ due to the cooperation of finite-size-effect,
Coulomb interaction and neutron-proton asymmetry. 
In addition, the surface energy of nucleus is analysed.
It is pointed out that a widely used phenomenological expression of
surface energy is not available in studying hot nucleus.
%

\newpage
  
\begin{center}
\begin{tabular}{|c|c|c|c|c|c|} \hline
 i &  $\mu_{i}$(fm) & $W_{i}$(MeV) & $B_{i}$(MeV) & $H_{i}$(MeV) &
 $M_{i}$(MeV)\\ \hline
 1 & 0.7& -402.4& -100 & -496.2 & -23.56 \\ \hline
 2 & 1.2 & -21.3 & -11.77 & 37.24 & -68.81\\ \hline
\multicolumn{6}{|c|}{ $t_0=1350$ MeV$\cdot$fm$^4$, $\alpha=1/3$,
$W_{LS}=115$ MeV$\cdot$fm$^5$, $X_0=1$.} \\ \hline
\end{tabular}
\end{center}
\begin{center}
Table 1. Parameters 
of Gogny D1 effective interaction.
\end{center}

\vspace {1cm}
\begin{center}
\begin{tabular}{|c|c|c|c|c|c|c|c|} \hline
\multicolumn{2}{|c|}{ }& $^{40}_{20}Ca$& $^{56}_{28}
Ni$ &$^{90}_{40}Zr$ &$^{156}_{62}Sm$
  &$^{208}_{82}Pb$ &$^{238}_{92}U$\\ \hline
  &Exp. &8.55 &8.64 &8.71 &8.25 &7.87 &7.57\\
  &Eq. (\ref{wncao}) &8.97 &8.82 &8.72 &8.02 &7.45 &7.15\\
Bound energy (MeV)  &Eq. (\ref{wnj})&-1.01 &-0.36 &0.66 &1.20 &1.30 &1.29\\
  &$E_{S}$   &8.77 &8.70 &8.70 &8.00 &7.41 &7.28\\
  &$E'_{S}$     &8.76 &8.72 &8.73 &8.11 &7.56 &7.27\\ \hline
 &Eq. (\ref{wncao})&0.131 &0.132 &0.134 &0.134 &0.134 &0.133\\
$\rho_{0}$(fm$^{-3}$) &$E_{S}$ &0.185 &0.180 &0.177 &0.173 &0.170 &0.176\\
 &$E'_{S}$ &0.129 &0.130 &0.133 &0.133 &0.133 &0.133\\ \hline
\end{tabular}
\end{center}
\begin{center}
Table 2. Zero-temperature saturation properties
of the six typical finite nuclei.
\end{center}

\newpage
\begin{center}
\begin{tabular}{|c|c|c|c|c|c|c|c|c|c|}\hline
\multicolumn{3}{|c|}{ } &$^{40}_{20}Ca$ &$^{56}_{28}Ni$ 
&$^{90}_{40}Zr$ &$^{152}_{62}Sm$
&$^{208}_{82}Pb$ &$^{238}_{92}U$ &SINM\\ \hline
(1)&$T_{c}(MeV)$ &Eq. (\ref{wncao}) &12.95 &13.15 &13.30 &13.35
&13.40 &13.35 &15.85\\
 & &Eq. (\ref{wnj}) &5.10 &6.35 &7.55 &8.50 &9.05 &9.20 &15.85\\ \hline
(1) &$\rho_{c}$(fm$^{-3}$) &Eq. (\ref{wncao}) &0.050 &0.050 &0.050
&0.055 &0.055 &0.060&0.060 \\
 & &Eq. (\ref{wnj}) &0.025 &0.030 &0.035 &0.040 &0.040 &0.040 &0.060\\ \hline
(2) &$T_{c}(MeV)$ &Eq. (\ref{wncao}) &12.35 &12.40 &12.45 &12.20 &12.05 &11.95&\\
& &Eq. (\ref{wnj}) &4.35 &5.35 &6.50 &7.30 &7.60 &7.65 & \\ \hline
(2) &$\rho_{c}$(fm$^{-3}$) &Eq. (\ref{wncao}) &0.050 &0.050 &0.050 &0.050
&0.050 &0.050 & \\
 & &Eq. (\ref{wnj}) &0.025 &0.030 &0.030 &0.035 &0.040 &0.040&\\ \hline
(2)&$T_{l}(MeV)$ &Eq. (\ref{wncao}) &9.45 &9.45 &9.45 &9.25 &9.10 &9.00&\\
 &$\rho_{l}$(fm$^{-3}$) &Eq. (\ref{wncao}) &0.075 &0.075 &0.075 &0.075 
&0.075 &0.075& \\ \hline
\end{tabular}
\end{center}
\begin{description}
\item {Table 3.} The limit temperature $T_{l}$,
density $\rho_{l}$ and liquid-gas phase transition critical
temperature $T_{c}$, density $\rho_{c}$ 
of six typical nuclei. (1) without
Coulomb interaction;
(2) with Coulomb interaction.
SINM=symmetric infinite nuclear matter.
\end{description}

\vspace{1cm}
\begin{center}
\begin{tabular}{|c|c|c|c|c|c|c|c|} \hline
\multicolumn{2}{|c|}{A} &50 &100 &200 &1000 &10000 &$\infty$\\ \hline
 & Eq. (\ref{wncao}) &13.10 &13.55 &13.95 &14.65 &15.30 & \\
$T_{c}(MeV)$ &$E'_{S}$ &12.65 &13.25 &13.80 &14.65 &15.30 &15.85 \\
 & Eq. (\ref{wnj}) &5.15 &8.05 &9.70 &12.25 &14.20 & \\ \hline
 &Eq. (\ref{wncao}) &0.050 &0.055 &0.055 &0.055 &0.060 & \\
$\rho_{c}$(fm$^{-3}$) &$E'_{S}$ & 0.050 &0.050 &0.055 &0.055 &0.060 &0.060 \\
 & Eq. (\ref{wnj}) &0.025 &0.035 &0.040 & 0.050 & 0.055 & \\ \hline
\end{tabular}
\end{center}
\begin{description}
\item{Table 4.} Size dependence of critical point
of liquid-gas transition
in symmetric infinite nuclear matter.
\end{description}

\newpage
\noindent
{\bf Figure Captions}
\begin {description}
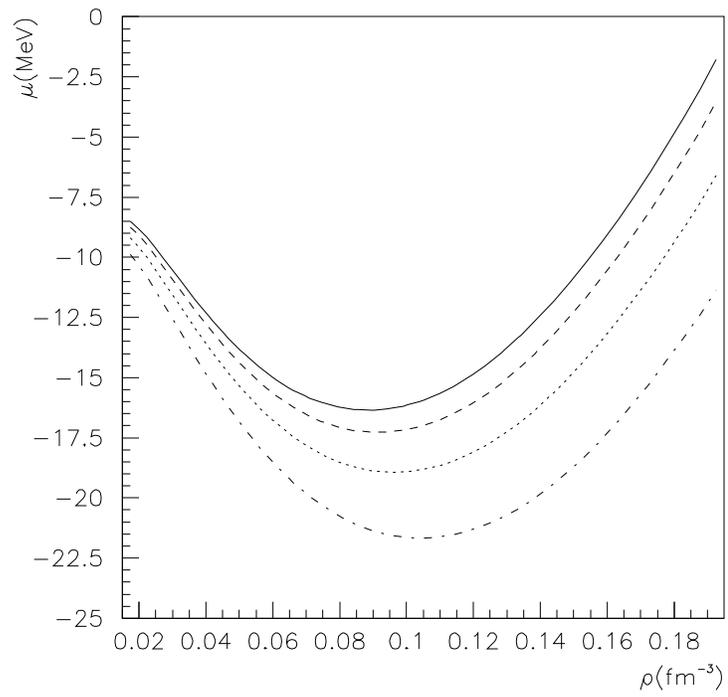
\item[Fig. 1]
The $\mu \sim \rho$ isotherms of infinite 
symmetric nuclear matter at $T=6.0$ MeV evaluated
with nucleon number $A=100$ (solid line), $A=200$ (dashed line),
$A=1000$ (dotted line) and $A=\infty$ (dash-dotted line) respectively.
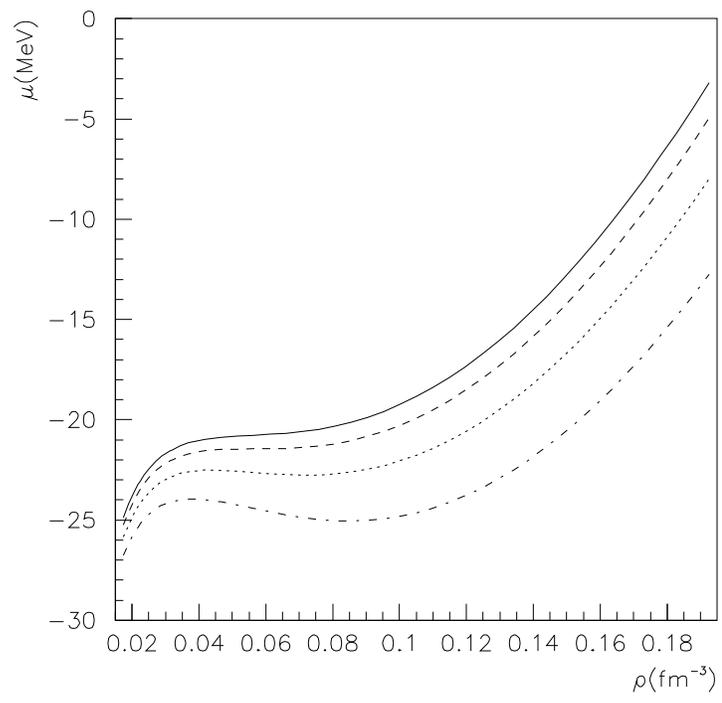
\item[Fig. 2]
Similar as Fig. 1 and $T=14.0$ MeV.
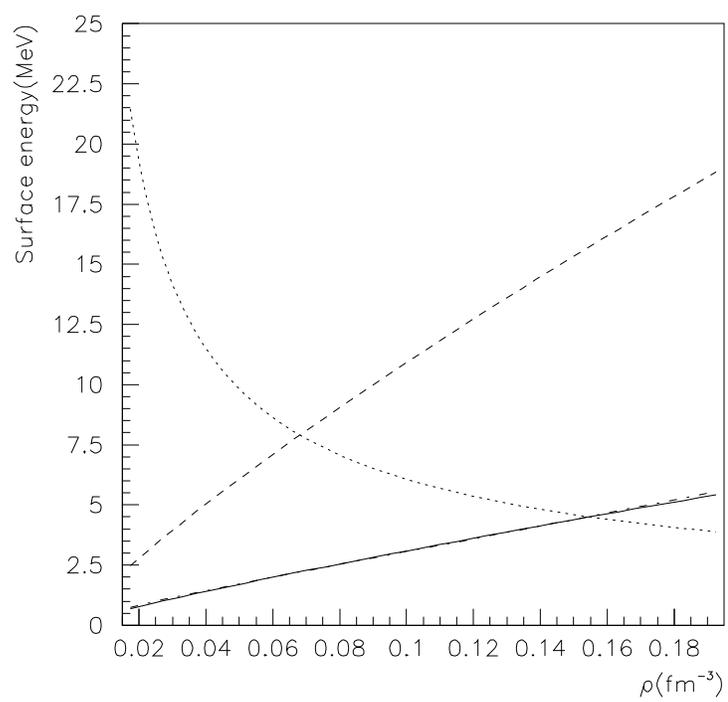
\item [Fig. 3]
Surface energy of $^{90}_{40}Zr$ at temperature 
$T=0.0$ MeV evaluated with Eqs. (\ref{sfc}) and (\ref{wncao}) with 
$a_{F}=0.35$ (solid line);
Eqs. (\ref{sfc}) and (\ref{wnj}) 
(dashed line); Eq. (\ref{sfp})
(dotted line); and
Eq. (\ref{sfphcao}) (dash-dotted line) respectively.  
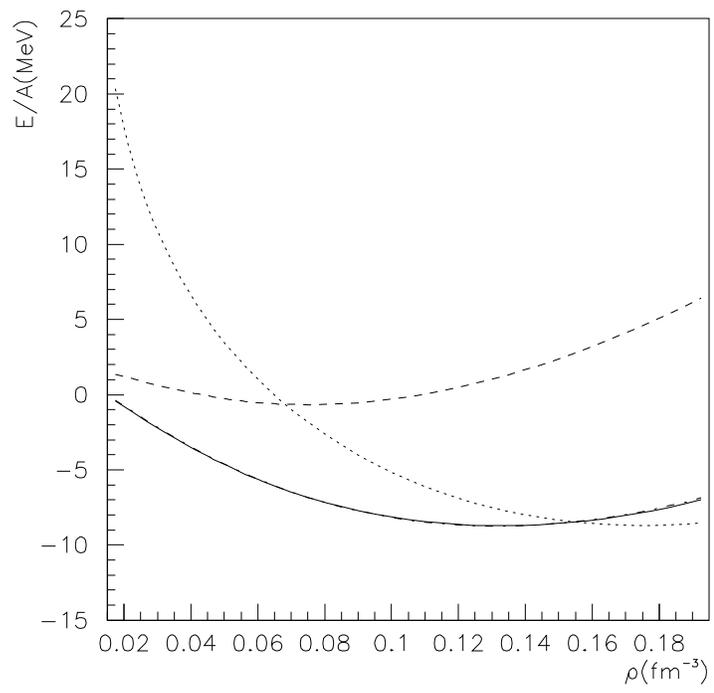
\item[Fig. 4]
The saturation curves $(E/A)\sim\rho$ calculated with Eq. (\ref{wncao})
(solid line), Eq. (\ref{wnj}) (dashed line),
Eq. (\ref{sfp})
(dotted line) and
Eq. (\ref{sfphcao}) (dash-dotted line) respectively.  
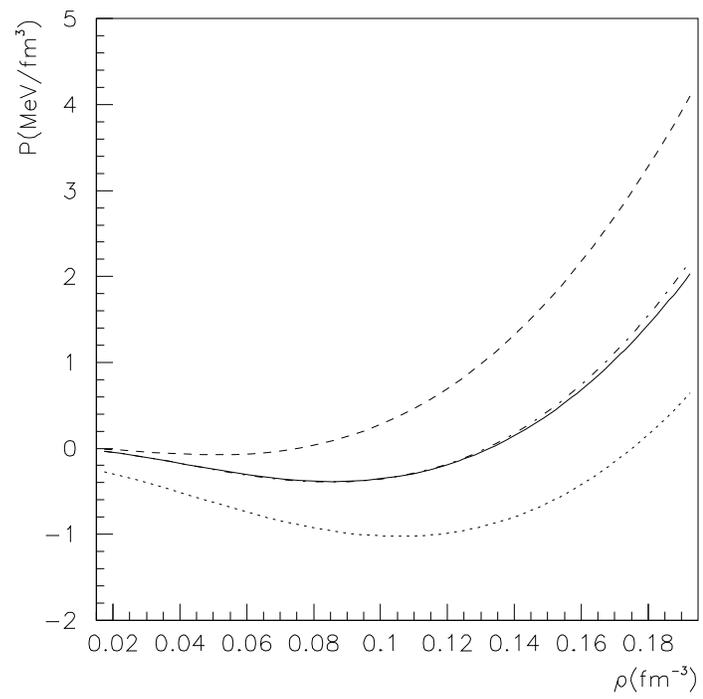
\item[Fig. 5]
The $P\sim \rho$ isotherms. The explanation of the curves is 
similar to Fig. 4.

\end {description}

\begin{thebibliography}{99}
\bibitem{Brack} Brack M, Guet C and Hakansson H B 1985 {\sl Phys. Rep.} 
{\bf 123} 277.
\bibitem{Li}Li G Q and Xu G O 1990 {\sl Phys. Rev.} C {\bf 42} 290.
\bibitem{Sauer}
Sauer G, Chandra H and Mosel U 1976 {\sl Nucl. Phys.} A {\bf 264} 221.
\bibitem{JMZ}
Jaqaman H R, Mekjian A Z and Zamick L 1983 {\sl Phys. Rev.} C {\bf 27} 2782.
\bibitem{Lej}
Lejenue A, Grange P, Martzollf M and Cugnon J 1986 {\sl Nucl. Phys.} A 
{\bf 453} 189.
\bibitem{Heyer}Heyer J, Kuo T T S, Shen J P and Wu S S 1988 
{\sl Phys. Lett.} B {\bf 202} 465.
\bibitem{Jiang}
Jinag M F, Heyer J, Yang S D and Kuo T T S 1988 {\sl Phys. Rev. Lett.} 
{\bf 61} 38.
\bibitem{Surk1}
Su R K, Yang S D and Kuo T T S 1987 {\sl Phys. Rew.} C {\bf 35} 1539.
\bibitem{Surk2}
Su R K, Li G Q and Kuo T T S 1986 {\sl Mod. Phys. Lett.} A {\bf 1} 71.
\bibitem{kuch}
Kucharek H, Ring P, Schuck P, Bengtsson R and Girod M 1989 
{\sl Phys. Lett.} B {\bf 216} 249.
\bibitem{Jaqaman}Jaqaman H R, Mekjian A Z and Zamick L 1984
	 {\sl Phys. Rev.}  C {\bf 29} 2067.
\bibitem{Song}Song H Q, Zheng G D and Su R K 1990
	{\sl J. Phys. G: Nucl. Part. Phys.} {\bf 16} 1861.
\bibitem{Huang}Huang S W, Fu M Z and Yang S D 1990 {\sl Mod. Phys. Lett.}
	  A {\bf 5} 1071.
\bibitem{Wang}Wang N P and Yang S D 1992 {\sl Acta Physica Sinica} {\bf 41}, 
	561 (in Chinese).
\bibitem{Gogny}
Decharge J and Gogny D 1980 {\sl Phys. Rew.} C {\bf 21} 1568.
\bibitem{Yang}Yang S D, Jiang M F and Heyer J 1989 {\sl Phys. Rev.} 
C {\bf 39} 2065.
\bibitem{Pan}Panagiotou A D, Curtin M W, Toki H, Scott D K,
	and Siemens P J 1984 {\sl Phys. Rev. Lett.} {\bf 52} 496.
\bibitem{Bondorf}Bondorf J, Donangelo R, Mishustin I N and Schulz H 1985
	{\sl Nucl. Phys.} A {\bf 444} 460.
\bibitem{Jaqaman2}Jaqaman H R 1989 {\sl Phys. Rev.} C {\bf 39} 169.
\bibitem{Band}Bandyopadhyay D, De J N and Skamaddar S K 1989
	{\sl Phys. Lett.} B {\bf 218} 391.
\end{thebibliography}
\end{document}